\documentclass[
  ,final            % use final for the camera ready runs
  ]
  {aipproc}

\layoutstyle{8x11single}

\usepackage{amssymb}
\usepackage{epsf}
\usepackage{color}
\usepackage{amsmath}
\usepackage{graphicx}
\begin{document}

\title{How to Remedy the $\eta$-Problem of SUSY GUT
Hybrid Inflation via Vector Backreaction}


\author{George Lazarides}{
  address={Physics Division, School of Technology, Aristotle University of 
  Thessaloniki, Thessaloniki 54124, Greece}
}

%\author{<author2>}{
  %address={<common address for author2 and author3>}
%}

%\author{<author3>}{
 % address={<common address for author2 and author3>}
  %,altaddress={<author1 address>} % additional visiting address
%}

\begin{abstract}

It is shown that, in supergravity models of inflation where the gauge 
kinetic function of a gauge field is modulated by the inflaton, we 
can obtain a new inflationary attractor solution, in which the 
roll-over of the inflaton suffers additional impedance due to the 
vector field backreaction. As a result, directions of the scalar 
potential which, due to strong K\"{a}hler corrections, become too 
steep and curved to normally support slow-roll inflation can now 
naturally do so. This solves the infamous $\eta$-problem of inflation 
in supergravity and also keeps the spectral index of the curvature 
perturbation mildly red despite $\eta$ of order unity. This mechanism 
is applied to a model of hybrid inflation in supergravity with a 
generic K\"{a}hler potential. The spectral index of the curvature 
perturbation is found to be 0.97 - 0.98, in excellent agreement with 
data. The gauge field can act as vector curvaton generating statistical 
anisotropy in the curvature perturbation. However, this anisotropy 
could be possibly observable only if the gauge coupling constant is
unnaturally small.
 
\end{abstract}

\maketitle

%%%%%%%%%%%%%%%%%%%%%%%%%%%%%%%%%%%%%%%%%%%%
%% MAINMATTER
%%%%%%%%%%%%%%%%%%%%%%%%%%%%%%%%%%%%%%%%%%%%

\section{Introduction}

The supersymmetric (SUSY) version \cite{susyhybrid} of hybrid 
inflation \cite{hybrid} is undoubtedly one of 
the most promising types of inflation models that remain into 
effect today. It naturally arises within SUSY grand unified 
theory (GUT) models and avoids the massive fine-tuning 
required in single field inflation models. The slowly rolling 
inflaton stays sub-Planckian during inflation and, thus, 
non-renormalizable terms are not totally out of control.
Inflation takes place at energies near the SUSY GUT scale in 
order to yield the observed magnitude of the curvature 
perturbation. Consequently, the waterfall field can be 
naturally identified with the GUT Higgs field.
SUSY is crucial for inflation since it offers a multitude of 
flat directions along which inflation can take place. Moreover, 
the flatness of these directions is not destroyed by radiative 
corrections as in non-SUSY theories. Actually, in SUSY hybrid 
inflation, the radiative corrections just provide a gentle 
logarithmic slope needed for the inflaton to slow-roll.

However, promoting global SUSY to local supergravity (SUGRA), 
the need of another set of fine-tunings is introduced. Indeed, 
K\"{a}hler corrections to the scalar potential generically 
give rise \cite{randall} to an inflaton mass of order the Hubble 
parameter $H$. This is the infamous $\eta$-problem of SUGRA 
inflation, i.e. the fact that the slow-roll parameter $\eta$ is 
pushed to order unity by SUGRA corrections:
\begin{equation}
\eta\equiv m_P^2\frac{V''}{V}\simeq\frac13
\left(\frac{m}{H}\right)^2={\cal O}(1)
\end{equation}
with \mbox{$m_P=2.44\times10^{18}~{\rm GeV}$} being the reduced 
Planck mass, \mbox{$m\sim H$} being the inflaton mass, and prime 
denoting derivative of the scalar potential $V$ with respect to the 
inflaton field. The scalar spectral index $n_s$ of the curvature 
perturbation then receives a contribution 
\begin{equation}
\delta(n_s-1)=2\eta={\cal O}(1),
\end{equation}
which contradicts its observed approximate scale invariance.
Moreover, SUGRA corrections lift the flatness of the inflaton 
direction and destabilize slow-roll. As a consequence, the 
number of e-foldings generated is not enough to solve the 
horizon and flatness problems of standard big bang cosmology.
Finally, if the inflaton mass is \mbox{$m\gtrsim \frac{3}{2}H$}, 
the inflaton cannot generate the necessary density perturbations.
Fortunately, hybrid inflation with a minimal K\"{a}hler 
potential is protected from SUGRA corrections by a cancellation  
and the $\eta$-problem does not arise \cite{sugrahybrid}.
However, any higher order corrections to the minimal 
K\"{a}hler potential would produce a massive $\eta$-problem.

Recently, a surprising solution to the $\eta$-problem was 
proposed \cite{Wagstaff}. An interaction of the inflaton field 
with a vector boson field leads to a new inflationary attractor 
solution, where the vector field backreaction ${\cal B}_A$ 
reduces the effective inflaton potential slope: 
\mbox{$|V'_{\rm eff}|<|V'|$} with \mbox{$V'_{\rm eff}\equiv 
V'+{\cal B}_A$}. This can overcome the $\eta$-problem by 
enabling long-lasting slow-roll inflation to take place even if 
$V$ is substantially steep and curved. Furthermore, the vector 
backreaction affects the inflaton equation of motion such that 
it allows the inflaton to undergo particle production even with 
\mbox{$\eta={\cal O}(1)$}. We show \cite{eta} that the mechanism 
of vector backreaction also 
protects the scalar spectral index $n_s$ against excessive 
contributions from a large $\eta$ parameter. Applying these 
results to the standard SUGRA hybrid inflation model with a 
generic non-minimal K\"{a}hler potential, we find \cite{eta} that 
hybrid inflation can be long-lasting and produce a weakly red 
spectrum of curvature perturbations in agreement with observations.

As expected, the vector field can, in principle, give rise 
\cite{anisinf,anisinf+} to statistical anisotropy in the curvature 
perturbation $\zeta$. Note that the observations still allow 
\cite{GE} as much as 30\% statistical anisotropy. Moreover, a 
preferred direction on the microwave sky might be hinted 
\cite{AoE} by the unlikely correlation of the low multiples of 
the cosmic microwave background. One way to generate statistical 
anisotropy is \cite{stanis} if the vector field acts as a 
curvaton \cite{sugravec,varkin}, i.e. via the vector field 
perturbations themselves. We apply \cite{eta} this to standard 
SUGRA hybrid inflation to see whether appreciable statistical 
anisotropy in the spectrum and bispectrum of the curvature 
perturbation can be generated. We consider natural units, where 
\mbox{$c=\hbar=k_B=1$}.

\section{Vector scaling slow-roll inflation}

Consider \cite{eta} a \mbox{$U(1)$} gauge symmetry with gauge 
field $B_{\mu}$ and a complex scalar Higgs field $\Phi$ with 
unit charge. Writing \mbox{$\Phi=\phi e^{i\varphi}/\sqrt2$} and 
using the gauge invariant combination \mbox{$h_0A_{\mu}\equiv 
h_0B_{\mu}-\partial_{\mu}\varphi$} (with $h_0$ being the gauge 
coupling constant), we obtain the Lagrangian density
\begin{equation}
 {\cal \mathcal{L}}=-\frac{1}{4}f(\sigma)F_{\mu\nu}F^{\mu\nu}
 +\frac12 \left(\partial_{\mu}\phi\right)\left(\partial^{\mu}
 \phi\right)
 +\frac{1}{2}h_{0}^{2}\phi^{2}A_{\mu}A^{\mu}
 -V_1(\phi),
\end{equation}
where $F_{\mu\nu}$ is the field strength and $V_1(\phi)$ is the  
potential for $\phi$. The gauge kinetic function $f$ is modulated 
by the real scalar inflaton field \mbox{$\sigma$}:
\begin{equation}
 f(\sigma)\equiv\left(\frac{h_{0}}{h(\sigma)}\right)^{2}
\end{equation}
with \mbox{$h(\sigma_0)=h_0$} so that \mbox{$f(\sigma_0)=1$} and 
$A_{\mu}$ becomes canonically normalized when $\sigma$ assumes its 
vacuum expectation value (VEV) $\sigma_0$. The spatial components 
of the physical vector field acquire a mass \mbox{$M_A=h_0\phi_0$} 
after $U(1)$ breaking ($\phi_0$ is the VEV of $\phi$).

As inflation homogenizes \mbox{$A_{\mu}$}, \mbox{$A_0$} vanishes 
\cite{vecurv} and, without loss of generality, we can take 
\mbox{$A_{\mu}=(0,0,0,A_z(t))$}. The metric is 
\mbox{$ds^2=dt^2-a_1^2(t)(dx^2+dy^2)-a_2^2(t)dz^2$}, where $a_{1,2}$ 
are the scale factors related to the different spatial directions. 
The average scale factor is \mbox{$a\equiv(a_1^2a_2)^{1/3}$} and the 
average Hubble rate is \mbox{$H\equiv\dot{a}/a$} (dot denotes 
derivative with respect to cosmic time). The anisotropic stress 
induced by the vector field is then
\begin{equation}
\Sigma\equiv\frac{1}{3H}\frac{d}{dt}\ln\left(\frac{a_1}{a_2}\right).
\end{equation}

The coupling of \mbox{$A_\mu$} to \mbox{$\sigma$} through the 
kinetic function \mbox{$f(\sigma)$} induces a source term 
\mbox{$\mathcal{B}_A\equiv - a_2^{-2}f'(\sigma)\dot{A}^2_z/2$} in 
the scalar field equation
\begin{equation}
\ddot{\sigma}+3H\dot{\sigma}+V'(\sigma)+\mathcal{B}_A(\sigma,
\dot{A}_z)=0,
\end{equation}
where prime denotes derivative with respect to \mbox{$\sigma$}, 
whose potential is $V(\sigma)$. For half of the parameter space 
and for flat $V(\sigma)$, the 
system evolves \cite{Wagstaff} to the standard slow-roll 
inflationary attractor. On this attractor \mbox{$\mathcal{B}_A$}, 
$\Sigma$, and the vector field energy density $\rho_A$ vanish and, 
thus, the vector field cannot influence the expansion of the 
universe. For the other half of the parameter space, 
\mbox{$\mathcal{B}_A$} backreacts \cite{Wagstaff} on the dynamics 
of $\sigma$ and the system tends to the vector scaling slow-roll 
(VSSR) inflationary attractor. On the VSSR attractor, $A_\mu$ has 
a non-negligible effect on the expansion of the universe via its 
non-vanishing $\rho_A$ and $\Sigma$. 

On this attractor, $f(\sigma)$ scales \cite{varkin} as 
\mbox{$f_{{\rm att}}\propto a^{-4}$}, which leads to 
scale-invariant transverse spectra \mbox{$\mathcal{P}_{L,R}$} 
of vector field perturbations. From parity conservation, the 
left and right transverse polarization components 
\mbox{$\mathcal{P}_{L,R}$} are identical. If the vector field 
is massless, its longitudinal component decouples and particle 
production of this field is highly anisotropic. Thus, the vector 
field, in this case, can only contribute subdominantly to $\zeta$, 
but it can still generate substantial statistical anisotropy and 
anisotropic non-Gaussianity.

The criteria for the existence of the VSSR attractor are 
\cite{Wagstaff}
\begin{equation}
 \textrm{Conditions\quad}
  \begin{cases}
    {\rm I}  \quad|\Gamma_{f}|\gg1,\\
    {\rm II}  \quad|\Gamma_{f}|\gg|\lambda_{0}|,\\
    {\rm III}  \quad\lambda_{0}\Gamma_{f}>6,
  \end{cases}
  \label{cond}
\end{equation}
where the dimensionless model parameters
\begin{equation}
\label{modelpara}
\Gamma_{f}(\sigma)\equiv\sqrt{\frac{3}{2}}m_{P}\left(\frac{f'}
{f}\right)
\quad{\rm and}\quad
\lambda_{0}(\sigma)\equiv\sqrt{\frac{3}{2}}m_{P}
\left(\frac{V'}{V}\right)\Biggr|_{\phi=0}
\end{equation}
are used and the vector field is considered massless. If the 
dimensionless model parameters are time-depended, we have 
additional conditions \cite{Wagstaff}:
\begin{equation}
\label{ABCD}
A,B,C,D<1,
\end{equation}
where 
\begin{eqnarray}
\nonumber
 A & \equiv & 4\sqrt{\frac{2}{3}}m_{P}\left|\frac{\Gamma_{f}'}{\Gamma_{f}^{2}}
 +\frac{1}{3}\frac{\lambda_{0}'}{\Gamma_{f}^{2}}\right|,\\ 
\nonumber
 B & \equiv & 2\sqrt{\frac{2}{3}}m_{P}\left|\frac{\lambda_{0}'}{\Gamma_{f}^{2}}\right|,\\
\nonumber
 C & \equiv & 4\sqrt{\frac{2}{3}}m_{P}\left|\frac{\Gamma_{f}'}{\Gamma_{f}^{2}}
 +\frac{\lambda_{0}'}{\Gamma_{f}^{2}}-\frac{\lambda_{0}'\Gamma_{f}
 +\lambda_{0}\Gamma_{f}'}{\Gamma_{f}\left(\lambda_{0}\Gamma_{f}-6\right)}\right|,\\
 D & \equiv & 4\sqrt{\frac{2}{3}}m_{P}\left|-\frac{2}{3}\frac{\lambda_{0}'}{\Gamma_{f}^{2}}
 +\left(\frac{\Gamma_{f}'}{\Gamma_{f}^{2}}\right)\left(\frac{2}{3}\frac{\lambda_{0}}{\Gamma_{f}}
 -\frac{8}{\Gamma_{f}^{2}}\right)\right|.
\end{eqnarray}

\subsection{General properties of the VSSR attractor}

The vector backreaction ${\cal B}_A\equiv - a_2^{-2}f'(\sigma)
\dot{A}^2_z/2$ is independent of $\dot{\sigma}$ and thus only 
modifies the effective slope of the potential 
\mbox{$V'_{\text{eff}}\equiv V'+\mathcal{B}_A$}. Since 
$f(\sigma)\propto 1/h^2\to 1$ after the end of inflation, we 
require that it is always decreasing in time, \mbox{$\dot{f}(t)<0$}, 
so that the $A_\mu$ remains weakly coupled during inflation. Then, 
since $\sigma$ decreases in time during inflation, 
\mbox{$f'(\sigma)>0$} and, thus, ${\cal B}_A<0$ thereby reducing the 
effective slope of the potential.
Once the dimensionless model parameters are slowly varying in time, 
the backreaction ${\cal B}_A$ is proportional to \mbox{$V'(\sigma)$} 
and we obtain \cite{Wagstaff} 
\begin{equation}
\label{effslope}
V'_{\textrm{eff}}\equiv V'+\mathcal{B}_A
\simeq\frac{6}{\lambda_{0}\Gamma_{f}}V'.
\end{equation}
Condition III in (\ref{cond}) then implies that the effective 
potential slope seen by the inflaton is reduced. So, we can 
obtain slow-roll inflation with potentials that would normally 
be too steep for slow-roll. Indeed, on the VSSR attractor, the 
slow-roll parameters \mbox{$\epsilon_{\textrm{H}}\equiv-
\dot{H}/H^2$} and \mbox{$\eta_{\textrm{H}}\equiv-\ddot{H}/2H
\dot{H}$} are
\begin{equation}
\label{slowroll}
 \epsilon_{\textrm{H}}\simeq\frac{2\lambda_{0}}{\Gamma_{f}}\ll1
 \quad\text{and}\quad
 \eta_{\textrm{H}}\simeq\frac{2\lambda_{0}}{\Gamma_{f}}
 +\frac{\sqrt{6}m_{P}}{\Gamma_{f}}\left(\frac{\lambda_{0}'}{\lambda_{0}}
 -\frac{\Gamma_{f}'}{\Gamma_{f}}\right)
\end{equation}
and slow-roll inflation with \mbox{$\epsilon_{\textrm{H}},
\eta_{\textrm{H}}\ll1$} is possible. From the slow-roll equations
\begin{equation}
\label{infeqs}
 3m_{P}^{2}H^{2}\simeq V(\sigma)\quad{\rm and}
 \quad3H\dot{\sigma}\simeq-V'_{\text{eff}}(\sigma),
\end{equation}
we can find the number of e-foldings
\begin{equation}
\label{efold}
 N_{\text{att}}=\frac{1}{4}\ln\frac{f(\sigma_{i})}
 {f(\sigma_{\text{end}})}
\end{equation}
as the inflaton rolls from \mbox{$\sigma_{i}$} to 
\mbox{$\sigma_{\text{end}}$}, its values at the start and the end of 
inflation, respectively.
The vector-to-scalar field energy density ratio $\mathcal{R}$ does 
not vanish \cite{Wagstaff} on the VSSR attractor:
\begin{equation}
\mathcal{R}\equiv\frac{\rho_{A}}{\rho_{\sigma}}
\simeq\frac{\lambda_{0}\Gamma_{f}-6}{\Gamma_{f}^{2}},
\end{equation}
inducing a small anisotropic stress \mbox{$\Sigma\simeq2\mathcal{R}/3$} 
leading \cite{anisinf,anisinf+} to statistical anisotropy in the 
curvature perturbation $\zeta$.

\subsection{Curvature perturbation from VSSR inflation}

At horizon crossing of the pivot scale \mbox{$k_*=0.002\text{Mpc}^{-1}$}, 
the curvature perturbation generated by the inflaton is \cite{eta} 
\begin{equation}
\label{zeta}
\frac{2}{5}\zeta_{\sigma}=\frac{\delta\rho_{\sigma}}{\rho_{\sigma}}
\Biggr|_{*}
=\frac{1}{5\sqrt{3}\pi}\frac{V^{3/2}}{m_{P}^{3}
\left|V'_{\textrm{eff}}\right|}\Biggr|_{*}
\simeq\frac{1}{30\sqrt{2}\pi}\frac{V^{1/2}|\Gamma_{f}|}{m_{P}^{2}}
\Biggr|_{*},
\end{equation}
where (\ref{effslope}) and (\ref{infeqs}) were used.
Considering that the observed \mbox{$\zeta\simeq4.8\times10^{-5}$} (COBE 
normalization \cite{wmap}) is generated by the inflaton alone, we have 
\mbox{$\zeta_{\sigma}\simeq\zeta$}. The spectrum of the curvature 
perturbation, in this case, is \cite{eta}
\begin{equation}
\mathcal{P}_{\zeta}\simeq
 \frac{1}{4\pi^{2}}\left(\frac{H^{2}}{\dot{\sigma}}\right)^{2}\Biggr|_{*}
 \simeq\frac{1}{24\pi^{2}m_{P}^{4}}\frac{V(\sigma)}{\epsilon(\sigma)}
 \left(\frac{\lambda_{0}\Gamma_{f}}{6}\right)^{2}\Biggr|_{*},
\end{equation}
where (\ref{infeqs}) was used and the slow-roll parameter $\epsilon$ is 
defined in the usual way $\epsilon(\sigma)\equiv (m_P^2/2)
\left(V'/V\right)^2$. The spectral index $n_{s}-1\equiv\frac{d\ln
\mathcal{P}_{\zeta}}{d\ln k}\Biggr|_{*}$
on the VSSR attractor is \cite{eta}
\begin{eqnarray}
\nonumber
 n_{s}-1 & \simeq & \left(\frac{6}{\lambda_{0}\Gamma_{f}}
 \right)\left[2\eta-6\epsilon-2m_{P}\sqrt{\frac{2}{3}}
 \frac{\left(\lambda_{0}\Gamma_{f}\right)'}{\Gamma_{f}}\right]\\
 & = & -2\left(\frac{6}{\lambda_{0}\Gamma_{f}}\right)
 \left[\epsilon+m_{P}\sqrt{\frac{2}{3}}\frac{\lambda_{0}
 \Gamma_{f}'}{\Gamma_{f}}\right],
 \label{eq: spectral index}
\end{eqnarray}
where $\eta$ is the usual slow-roll parameter \mbox{$\eta(\sigma)
\equiv m_P^2\left(V''/V\right)$}. The standard result \mbox{$n_{s}-1
=2\eta-6\epsilon$} is recovered for \mbox{$V'_{\text{eff}}=V'$}, i.e. 
\mbox{$\lambda_{0}\Gamma_{f}=6$}, hence \mbox{$\left(\lambda_{0}
\Gamma_{f}\right)'=0$}. We see that $n_s$ is independent of the 
potential curvature encoded in $\eta$ and it is easy to obtain a red 
spectrum favored by the current observational bounds 
\mbox{$0.953\leq n_s\leq0.981$} (at 1$\sigma$) \cite{wmap}.
The running of the spectral index
\mbox{$n_{s}'\equiv\frac{dn_{s}}{d\ln k}$} on the VSSR attractor is also 
found \cite{eta} to be
\begin{equation}
\label{running}
n_{s}'\simeq2\epsilon\left(\frac{6}{\lambda_{0}\Gamma_{f}}\right)^{2}
\left\{ \eta-2\epsilon+2m_{P}^{2}\left[\frac{\Gamma_{f}''}{\Gamma_{f}}
 -2\left(\frac{\Gamma_{f}'}{\Gamma_{f}}\right)^{2}\right]\right\}
\end{equation}
and should be compared with the current observational bounds \cite{wmap} 
with no gravitational waves: 
\mbox{$-0.084 < n_s' < 0.020$} (at \mbox{$95\%\text{cf}$}).
The tensor-to-scalar ratio on the VSSR attractor is \cite{eta}
\begin{equation}
\label{tensor}
r\simeq16\epsilon\left(\frac{6}{\lambda_{0}\Gamma_{f}}\right)^{2}=
\frac{192}{\Gamma_{f}^{2}}
\end{equation}
with current observational bound \cite{wmap} with no running of the
spectral index \mbox{$r < 0.36$} (at \mbox{$95\%\text{cf}$}).

\subsection{Statistical anisotropy and anisotropic non-Gaussianity}

Since, during VSSR inflation, the vector field remains massless but 
with a small non-zero energy density, it could contribute 
\cite{sugravec,vecurv} to the
primordial curvature perturbation $\zeta$. After the end of inflation, 
the vector field becomes heavy with mass $M_A$ and oscillates rapidly 
behaving like pressureless matter. It can then nearly dominate the 
energy density and imprint \cite{vecurv} its spectra of perturbations
generated during inflation. If \mbox{$f\propto a^{-4}$}, these spectra 
are \cite{varkin} 
\begin{equation}
\mathcal{P}_{L,R}=\mathcal{P}_{\sigma}
=\left(\frac{H_{*}}{2\pi}\right)^{2}\quad\text{and}\quad
\mathcal{P}_{\parallel}=0.
\end{equation}
The decay rate $\Gamma_{A}$ of the oscillating massive vector field is
\begin{equation}
\label{rate}
\Gamma_{A}=\frac{h_{0}^{2}M_{A}}{8\pi}=\frac{h_{0}^{3}\phi_{0}}
{8\pi}=H_{\text{dec}},
\end{equation}
where the subscript `dec' denotes the epoch of vector field decay.
For the gravitational effect of the vector field not to be suppressed, 
its oscillations should last \cite{eta} at least one Hubble time before 
its decay, 
i.e.
\begin{equation}
\label{varepsilon}
\tau\equiv\frac{\Gamma_{A}}{H_{\text{end}}}
\simeq \frac{h_0^3\phi_0|\Gamma_{f}(\sigma_{*})|}
{32\sqrt{6}\pi^2m_{P}\zeta}\lesssim1,
\end{equation}
where (\ref{zeta}) and (\ref{rate}) were used and the subscript `end' 
marks the end of inflation. From this, we get
\begin{equation}
\label{h0}
h_{0}\lesssim\left(\frac{32\sqrt{6}\pi^{2}m_{P}\zeta}
{\phi_0|\Gamma_{f}(\sigma_{*})|}
\right)^{1/3}.
\end{equation}

The anisotropic spectra of perturbations of $A_\mu$ generate statistical 
anisotropy parametrized by $g$:
\begin{equation}
\mathcal{P}_{\zeta}(\mathbf{k})
=\mathcal{P}^{\text{iso}}_{\zeta}(k)
\left[ 1+g\left(\mathbf{\hat{d}\cdot\hat{k}}\right)^2+\cdots \right],
\end{equation}
where \mbox{$\mathbf{\hat k}\equiv \mathbf k/k$} and $\mathbf{\hat d}$ 
is the unit vector in the preferred direction. The present upper bound 
on \mbox{$\left|g\right|$} is 0.3 \cite{GE}, while the Planck satellite 
will reduce \cite{planck} it to \mbox{$0.02$}. Maximal statistical 
anisotropy is generated \cite{eta} if the inflaton decays rapidly after 
inflation, while the vector curvaton decays later. In this optimal case, 
the statistical anisotropy is found to be \cite{eta} 
\begin{equation}
\label{aniso}
 \left|g\right|
 \approx\sqrt{\frac23}
 \frac{\phi_0\mathcal{R}_{\text{end}}}{m_Ph_0\zeta|\Gamma_f(\sigma_*)|}.
\end{equation}
For a Planck detectable statistical anisotropy \mbox{$\left|g\right|
\gtrsim0.02$} \cite{planck}, this gives
\begin{equation}
 h_{0}\lesssim 50
\sqrt{\frac{2}{3}}
 \frac{\phi_0\mathcal{R}_{\text{end}}}{m_{P}
 \zeta|\Gamma_{f}(\sigma_{*})|}.
\end{equation}

The vector curvaton may also generate non-Gaussianity in the 
curvature perturbation $\zeta$.
At present, there is \cite{wmap} a hint for a non-zero non-linearity 
parameter $f_{\text{NL}}$ characterizing non-Gaussianity: 
\mbox{$f_{\text{NL}}^{local}=32\pm21$} (at 1$\sigma$) .
In the optimal case discussed above, we obtain \cite{eta}
\begin{equation}
\label{fNL}
f_{\text{NL}}\simeq\frac{5}{3}g^{2}\frac{\sqrt{\tau}}
{\mathcal{R}_{\text{end}}}.
\end{equation}
For Planck detectable non-Gaussianity \mbox{$f_{\text{NL}}\gtrsim
\mathcal{O}(1)$}, this leads to
\begin{equation}
 h_{0}
 \lesssim\frac{25\sqrt{6}}{3888}\frac{\mathcal{R}_{\text{end}}^{2}}{\pi^{2}
 |\Gamma_{f}^{3}(\sigma_{*})|}\left(\frac{\phi_0}{m_{P}\zeta}\right)^{5}.
\end{equation}

\section{Vector scaling SUGRA hybrid inflation}

We now embed \cite{eta} the above vector curvaton into a well motivated 
model of SUSY GUT hybrid inflation. Consider a simple SUSY GUT model based on 
the gauge group  \mbox{$G=G_{\rm SM}\times U(1)_{B-L}$} with $G_{\rm SM}$
being the standard model gauge group, which naturally incorporates 
\cite{susyhybrid,sugrahybrid} standard SUSY hybrid inflation. The gauge 
group $G$ may be thought as part of a larger GUT gauge symmetry (see 
e.g. \cite{peddie}). The model contains a conjugate pair of $G_{\rm SM}$ 
singlet superfields $\Phi$ and $\bar{{\Phi}}$ with $B-L=+1$ and $-1$, 
respectively, which 
break \mbox{$U(1)_{B-L}$} by their VEVs and a gauge singlet $S$ triggering 
the \mbox{$U(1)_{B-L}$} breaking and acting as our slowly rolling inflaton.
Adequate flatness of the inflationary trajectory is guaranteed by a 
discrete $Z_n$ R-symmetry: \mbox{$S\rightarrow Se^{2\pi i/n}$}, 
\mbox{$W\rightarrow We^{2\pi i/n}$}. Note, in passing, that such 
symmetries arise 
\cite{classification} in many compactified string theories and can 
effectively act \cite{stringinspired} as continuous symmetries.

The most general superpotential relevant for inflation and allowed by the 
symmetries of the model is
\begin{equation}
 W= S\sum_{k_1,k_2=0}^{\infty}A_{k_1k_2}\left(\Phi\bar{{\Phi}}\right)^{k_1}
 \left(S^n\right)^{k_2},
\end{equation}
where \mbox{$A_{k_1k_2}$} are coefficients with varying dimensions (for a 
similar analysis, see \cite{tetradis}). For $n\geq3$, we rewrite this 
as
\begin{equation}
W=\kappa S\left(\Phi\bar{{\Phi}}-M^{2}\right) + \text{``non-renormalizable terms''},
\end{equation}
where \mbox{$A_{00}=-\kappa M^2$} and \mbox{$A_{10}=\kappa$} with $\kappa$ and
$M\simeq M_{\rm GUT}$ made positive by field redefinitions ($M_{\rm GUT}$ is the 
SUSY GUT scale). The non-renormalizable terms are suppressed by powers of $m_P$. 

The scalar potential in SUGRA has the form
\begin{equation}
V=e^{K/m_P^2}\left[F_{\Phi_i}K^{-1}_{ij*}F_{\Phi^*_j} -3\frac{|W|^2}{m_P^2} \right]
+\frac12 \sum_{a,b}\left[\text{Re}f_{ab}(\Phi_i)\right]^{-1}h_ah_bD_aD_b,
\label{Vsugra}
\end{equation}
where $K$ is the K\"{a}hler potential, $f_{ab}$ the gauge kinetic functions, 
and  
\begin{equation}
K_{ij*}=\frac{\partial^2K}{\partial\Phi_i\partial\Phi^*_j}\,,\quad
F_{\Phi_i}=\frac{\partial W}{\partial \Phi_i}+
\frac{W}{m_P^2}\frac{\partial K}{\partial \Phi_i}\,,\quad
D_a=\Phi_i(T_a)^i_j\frac{\partial K}{\partial \Phi_j}+\xi_a.
\end{equation}
Here the subscripts \mbox{$a,b,\cdots$} label the generators $T_a$ of the gauge 
group with gauge couplings $h_a$ and $\xi_a$ are Fayet-Iliopoulos D-terms for 
the $U(1)$ gauge groups. In our model, only the gauge kinetic function $f$ for 
the $U(1)_{B-L}$ is taken not equal to unity and we have just three fields 
\mbox{$\Phi_i=(S,\Phi,\bar{{\Phi}})$}. D-flatness requires that 
\mbox{$\bar{\Phi}^*=\Phi e^{i\theta}$}, where we choose $\theta=0$ so that the 
SUSY vacua are contained in this D-flat direction. Then, bringing $\Phi$, 
$\bar{{\Phi}}$ on the real axis by a $U(1)_{B-L}$ rotation, we write 
\mbox{$\Phi=\bar{\Phi}\equiv\phi/2$}, where $\phi$ is a normalized real scalar 
field. We also define the normalized real scalar field $\sigma$: 
\mbox{$|\sigma|\equiv\sqrt{2}|S|$} (see below). The SUSY minimum is at 
\mbox{$\sigma=\sigma_0=0$} and \mbox{$\phi=\phi_0=\pm2M$}.

For \mbox{$|\sigma|>|\sigma_c|=\sqrt{2}M$}, the potential $V$ in global SUSY 
has a stable flat direction at \mbox{$\phi=0$} along which inflation can 
take place driven by the false vacuum energy density \mbox{$\kappa^2M^4$}. As 
$|\sigma|$ crosses its critical value $|\sigma_c|$, the effective 
mass-squared $m_{\phi}^{2}$ of \mbox{$\phi$}
\begin{equation}
m_{\phi}^{2}\simeq\kappa^{2}\left(|S|^{2}-M^{2}\right)
\end{equation}
becomes tachyonic and inflation ends abruptly by a waterfall.

The K\"{a}hler potential $K$ is a real function of the invariants 
\mbox{$|S|^2,|\Phi|^2,|\bar{\Phi}|^2,\Phi\bar{\Phi}$}, $S^n$. On the 
inflationary trajectory, where \mbox{$\Phi=\bar{\Phi}=0$}, the matrix 
\mbox{$K_{ij*}$} becomes diagonal and \mbox{$F_{\Phi}=F_{\bar{\Phi}}=0$}. 
So the only terms in $K$ which will contribute on this trajectory are
\begin{equation}
 K= \sum_{k_1,k_2=0}^{\infty} \frac{|S|^{2k_1}}{m_P^{2k_1+nk_2-2}} 
 \left[a_{k_1k_2}\left(S^n\right)^{k_2}+h.c\right],
\end{equation}
where \mbox{$a_{k_1k_2}$} are dimensionless coefficients of order 
unity. Hence \mbox{$K=|S|^2-(\alpha/4)|S|^4/m_P^2+\cdots$} with 
\mbox{$a_{00}=0$}, \mbox{$a_{10}=1/2$}, and \mbox{$a_{20}=-\alpha/8$}, 
where \mbox{$|\alpha|\sim1$} is a real parameter.

The SUGRA scalar potential in (\ref{Vsugra}) restricted on the 
inflationary trajectory can then be parameterized as
\begin{equation}
V=\kappa^2M^4\sum_{k_1,k_2=0}^{\infty} 
P_{k_1k_2}\frac{|S|^{2k_1}(S^n)^{k_2}}{m_P^{2k_1+nk_2}}+h.c.,
\end{equation}
where the \mbox{$P_{k_1k_2}$} are dimensionless coefficients, which 
are functions of \mbox{$A_{k_1k_2}$} and \mbox{$a_{k_1k_2}$}. 
Writing \mbox{$S=|\sigma|e^{i\vartheta}/\sqrt2$}, the dimensionless 
inflationary potential $V/\kappa^{2}M^{4}$ is found \cite{eta} to be 
\begin{equation}
\label{dimpot}
\frac{V}{\kappa^{2}M^{4}}=1+\frac{\alpha}{2}\left(\frac{\sigma}{m_{P}}\right)^{2}
 +\beta\left(\frac{\sigma}{m_{P}}\right)^{4}+
 2\gamma(n+1)\left(\frac{|\sigma|}{\sqrt2m_P}\right)^n\cos n\vartheta
 +\cdots,
\end{equation}
where $8\beta=1+7\alpha/2+2\alpha^2-18(a_{30}+c.c)$, $\gamma=c+a_{01}-a_{11}$ 
taken real with $c$ from \mbox{$A_{01}=-c\kappa M^2/m_P^n$}. Minimizing $V$ 
with respect to $\vartheta$, we find that, for \mbox{$\gamma<0$}, 
\mbox{$\vartheta=2\pi k/n$}, which by a $Z_n$ transformation can be brought to 
zero. For \mbox{$\gamma>0$}, $V$ is minimized with \mbox{$\vartheta=(2k+1)\pi/n$}, 
which by a $Z_n$ transformation can be brought to \mbox{$\pi/n$}.

The one-loop radiative corrections to the inflationary potential are 
given by the Coleman-Weinberg formula \cite{cw}:
\begin{equation}
\Delta V_{\rm 1-loop}=\frac{\left(\kappa M\right)^{4}}{32\pi^{2}}
\left(2\ln\frac{\kappa^{2}\sigma^{2}}{2Q^{2}}+f_c(x)\right)
\end{equation}
with \mbox{$f_c(x)\equiv (x+1)^2\ln (1+1/x)+(x-1)^2\ln (1-1/x)$}, where 
\mbox{$x\equiv \sigma^2/2M^2$} and $Q$ is a renormalization scale.
They generate a logarithmic slope on the inflationary valley.
Note that, for sub-Planckian field values \mbox{$\sigma< m_{P}$}, the 
scalar fields remain approximately canonically normalized even with a 
non-minimal K\"{a}hler potential. As we see from (\ref{dimpot}), 
for non-canonical K\"{a}hler potential $K$, the inflaton obtains a 
contribution to its mass-squared \mbox{$V''(\phi=0)\simeq3\alpha H^{2}$} 
leading to \mbox{$\eta\simeq\alpha$}. Thus, $\sigma$ would normally 
be fast-rolling unless \mbox{$\alpha$} is suppressed, which is the 
infamous $\eta$-problem. In the VSSR attractor, however, we can obtain 
slow-roll inflation even without fine-tuning the non-canonical K\"{a}hler 
coefficient $\alpha$.

In our case, the dimensionless model parameter $\lambda_{0}$ in 
(\ref{modelpara}) becomes \cite{eta}
\begin{equation}
\label{lambda0}
\lambda_{0}=\sqrt{\frac{3}{2}}\left[\alpha\left(\frac{\sigma}{m_{P}}\right)
+\left(4\beta-\frac{1}{2}\alpha^{2}\right)\left(\frac{\sigma}{m_{P}}\right)^{3}
 +\frac{\kappa^{2}}{8\pi^{2}}\left(\frac{m_{P}}{\sigma}\right)+\cdots\right].
\end{equation}
We assume here natural values for \mbox{$|\alpha|\sim1$} so that the first 
term in $\lambda_0$ dominates over the radiative corrections until the end of 
inflation. The condition for this is 
$\kappa<4\pi\sqrt{|\alpha|}\left(M/m_{P}\right)$. We will show that a 
red spectrum can still be obtained, in this case, if the cosmological scales 
exit during the VSSR attractor. We also find for the first derivative 
of $\lambda_{0}$ that 
\begin{equation}
\sqrt{\frac{2}{3}}m_{P}\left|\lambda_{0}'\right|
=\left|\eta-2\epsilon\right|\simeq
\Big|\alpha +3\left(4\beta-\frac{1}{2}\alpha^{2}\right)
\left(\frac{\sigma}{m_{P}}\right)^{2}+\cdots \Big|.
\end{equation}

\subsection{An exponential gauge kinetic function}

The gauge kinetic function, which is holomorphic, cannot contain terms 
linear in $S$ because of the $Z_n$ R-symmetry, but combinations $S^n$ are 
allowed. Combinations $\Phi\bar{\Phi}$ are also allowed, but they do not 
contribute on the inflationary trajectory. We consider \cite{eta} an 
exponential gauge kinetic function
\begin{equation}
\label{kinetic}
 f(S^n)=\exp\left[q\left(\frac SM\right)^n\right]
 =\exp\left[q\left(\frac{|\sigma|}{\sqrt2M}\right)^ne^{in\vartheta}\right],
\end{equation}
which goes to unity as the inflaton settles into the SUSY vacuum.
For \mbox{$\gamma<0$}, where \mbox{$\vartheta=0$}, we choose $q>0$ and for 
\mbox{$\gamma>0$}, where \mbox{$\vartheta=\pi/n$}, we choose $q<0$ so that 
the exponent in (\ref{kinetic}) is always positive:
\begin{equation}
f(\sigma)=e^{|q|\left(\frac{|\sigma|}{\sqrt 2M}\right)^n}.
\end{equation}

The dimensionless model parameter $\Gamma_{f}(\sigma)$ in (\ref{modelpara}) 
is then
\begin{equation}
\Gamma_{f}(\sigma)=|q|n{\frac{\sqrt 3}{2}}\left(\frac{m_{P}}{M}\right)
\left(\frac{\sigma}{|\sigma|}\right)
\left(\frac{|\sigma|}{\sqrt 2M}\right)^{n-1}.
\end{equation}
We see that this parameter is not constant for \mbox{$n\neq1$.}
The VSSR attractor demands that $\Gamma_f$ and $\lambda_0$ have the 
same sign, so we take the non-canonical K\"{a}hler parameter $\alpha>0$. 
We also choose \mbox{$\sigma>0$} for definiteness.
For \mbox{$|q|\sim1$}, conditions I and II in (\ref{cond}) for the 
existence of the VSSR attractor are readily satisfied, while condition 
III requires that $|q|\alpha n>4$. Since \mbox{$\lambda_0=\lambda_0(\sigma)$} 
and \mbox{$\Gamma_f=\Gamma_f(\sigma)$} are not constants, we have to consider 
the additional conditions in (\ref{ABCD}) too. The tightest ones are 
\mbox{$A,C<1$}, which yield 
\begin{equation}
|q|\gtrsim\frac83-\frac{8}{3n}, \quad 
\alpha\gtrsim 4\left[n\left(|q|-\frac83\right)\right]^{-1}
\end{equation}
with the latter being stronger than $|q|\alpha n>4$ required by 
condition III in (\ref{cond}).

\subsection{Properties of vector scaling slow-roll inflation}

The e-foldings $N_{\text{att}}$ during the VSSR attractor from an
initial inflaton value $\sigma_i$ until the end of inflation are
found from (\ref{efold}) to be
\begin{equation}
N_{\text{att}}=\frac{|q|}{4}\left[\left(\frac{\sigma_i}
{\sqrt 2M}\right)^n-1\right].
\end{equation}
For \mbox{$\sigma_i<m_P$} and, say, \mbox{$|q|=n=3$}, we find that 
\mbox{$N_{\text{att}}\lesssim10^5$}, which is more than enough for 
solving the horizon and flatness problems. At horizon exit of the 
pivot scale $k_*$,
\begin{equation}
\label{sigma*}
\frac{\sigma_{*}}{\sqrt2M}=\left(\frac{4 N_{*}}{|q|}+1\right)^{1/n}.
\end{equation}
The reduced effective scalar potential slope is found from 
(\ref{effslope}) to be given by
\begin{equation}
\frac{V'_{\textrm{eff}}}{V'}\simeq\frac{4}{|q|\alpha n}\left(
\frac{\sqrt 2M}{\sigma}\right)^n\ll1.
\end{equation}
The slow-roll parameters in (\ref{slowroll}) become
\begin{equation}
\epsilon_{\textrm{H}}\simeq\frac{2\alpha}{|q|n}
\left(\frac{\sigma}{m_{P}}\right)^2\left(\frac{\sqrt 2M}{\sigma}\right)^n,\quad
\eta_{\textrm{H}}\simeq\epsilon_{\textrm{H}}+\frac{2(2-n)}{|q|n}
\left(\frac{\sqrt 2M}{\sigma}\right)^n
\end{equation}
and remain $\ll1$ for \mbox{$n\geq3$} and \mbox{$m_P>\sigma>\sigma_c=\sqrt2M$}.
The vector-to-scalar energy density ratio $\mathcal{R}$ increases
during VSSR inflation and its value at the end of inflation is 
\begin{equation}
\label{vectortoscalar}
\mathcal{R}_{\text{end}}\simeq 2\left(\frac{M}{m_{P}}\right)^{2}\left[\frac{|q|
\alpha n-4}{(qn)^{2}}\right].
\end{equation}
With \mbox{$n=|q|=3$}, \mbox{$\alpha=4$}, we find that 
\mbox{$\mathcal{R}_{\text{end}}\simeq 1\times10^{-4}$}. For large $n$, the 
ratio decreases as \mbox{$\mathcal{R}_{\text{end}}\propto1/n$}.

\subsection{The curvature perturbation}

The primordial curvature perturbation is found \cite{eta} from (\ref{zeta})
and (\ref{sigma*}) to be
\begin{equation}
\frac{2}{5}\zeta\simeq
\frac{\kappa |q|n}{20\sqrt{6}\pi}\left(\frac{M}{m_{P}}\right)
\left(\frac{4N_*}{|q|}+1\right)^{(n-1)/n}.
\end{equation}
The COBE normalization \cite{wmap} with \mbox{$M=M_{\rm GUT}$}, \mbox{$N_*=60$}, 
\mbox{$n\geq3$}, \mbox{$|q|\gtrsim3$} then leads to 
\mbox{$\kappa \lesssim1.5\times10^{-3}$}, which readily satisfies the 
condition for the radiative corrections to be subdominant in 
(\ref{lambda0}) with \mbox{$\alpha=4$}. Note that here the COBE 
normalization 
can be satisfied for \mbox{$M=M_{\rm GUT}$}, whereas, in standard SUSY 
hybrid inflation, $M$ turns out to be \cite{susyhybrid,sugrahybrid} 
somewhat below $M_{\rm GUT}$.
For \mbox{$|q|,\alpha\sim1$}, the scalar spectral index is found 
from (\ref{lambda0}), (\ref{kinetic}), and (\ref{sigma*}) to be  
\begin{equation}
n_{s}\simeq1-\frac{2(n-1)}{nN_*}.
\end{equation}
So, for \mbox{$n=3$}, \mbox{$n_s\simeq0.978$} and, for \mbox{$n\gg1$}, 
\mbox{$n_s\simeq0.967$}, which fit very well within the $1\sigma$ bounds 
from WMAP \cite{wmap}. Using (\ref{running}) and (\ref{lambda0}), we 
find that the running of $n_{s}$ in our model is well approximated by
\begin{equation}
n_{s}'\simeq-\frac{2(n-1)}{nN_*^2},
\end{equation}
which, for \mbox{$n=3$}, gives $n_s'\simeq-3.6\times10^{-4}$ and, for 
\mbox{$n\gg1$}, \mbox{$n_s'\simeq-5.4\times10^{-4}$} well satisfying 
the WMAP constraints \cite{wmap}. The tensor-to-scalar ratio is found
from (\ref{tensor}) and (\ref{lambda0}) to be
\begin{equation}
 r\simeq\frac{256}{(qn)^{2}}\left(\frac{M}{m_{P}}\right)^{2}
 \left(\frac{4N_*}{|q|}+1\right)^{2(1-n)/n},
\end{equation}
which, for \mbox{$N_*=60$}, \mbox{$n\geq3$}, \mbox{$|q|\gtrsim3$}, gives 
\mbox{$r\lesssim1\times10^{-6}$}, compatible with WMAP, but 
probably too small to be observed.

\subsection{Statistical anisotropy}

For the gravitational effect of the vector curvaton not to be suppressed
\mbox{$\tau\lesssim1$} with $\tau$ in (\ref{varepsilon}). 
This requirement, using (\ref{h0}) and (\ref{sigma*}), gives
\begin{equation}
h_{0}^3\lesssim\frac{32\sqrt2\pi^{2}\zeta}{|q|n}
\left(\frac{4N_*}{|q|}+1\right)^{(1-n)/n},
\end{equation}
which, for \mbox{$N_*=60$}, \mbox{$n\geq3$}, \mbox{$|q|\gtrsim3$}, 
becomes \mbox{$h_0\lesssim 0.05$}. So, for the SUSY GUT value 
\mbox{$h_0\sim0.7$} of the gauge coupling constant, the  gravitational 
effect of the vector curvaton is somewhat suppressed. For the 
statistical anisotropy induced in the spectrum (see (\ref{aniso}))
\begin{equation}
\left|g\right|\approx\frac{4\sqrt2\mathcal{R}_{\text{end}}}{3|q|nh_0\zeta}
\left(\frac{M}{m_P}\right)^{2}\left(\frac{4N_*}{|q|}+1\right)^{(1-n)/n},
\end{equation}
where $\mathcal{R}_{\text{end}}$ is given in (\ref{vectortoscalar}), to 
be Planck detectable \cite{planck}, \mbox{$h_0\lesssim1\times10^{-4}$} for 
$\alpha\simeq4$, \mbox{$N_*=60$}, \mbox{$n\geq3$}, \mbox{$|q|\gtrsim3$}, 
which is far smaller than the SUSY GUT $h_0$. The amplitude of non-Gaussianity
(see (\ref{fNL}))
\begin{equation}
 f_{\text{NL}}\simeq\frac53g^2\frac{\sqrt{\tau}}{\mathcal{R}_{\text{end}}}
\end{equation}
is Planck detectable, i.e. \mbox{$f_{\text{NL}}\gtrsim\mathcal{O}(1)$}, if 
\mbox{$h_0\lesssim3\times10^{-10}$} for \mbox{$\alpha\simeq4$}, \mbox{$N_*=60$}, 
\mbox{$n\geq3$}, \mbox{$|q|\gtrsim3$}, which is far too small.

\section{Conclusions}

We showed that, if the inflaton modulates the kinetic function of a 
vector field, the backreaction to the inflaton's variation allows 
slow-roll inflation despite the fact that the inflationary potential 
is too steep as a result of sizable K\"{a}hler corrections. Moreover,
a mildly red spectral index of inflaton perturbations can be produced
eliminating the $\eta$-problem of SUGRA inflation.

We have applied the above mechanism to a model of SUGRA hybrid 
inflation, where the waterfall field is taken to be the Higgs field 
of a SUSY GUT. The vector field with modulated kinetic function is 
one of the GUT gauge bosons which become massive at the GUT phase 
transition. We showed that slow-roll inflation can take place with a 
generic K\"{a}hler potential despite the fact that 
\mbox{$\eta={\cal O}(1)$}. Moreover, a red spectrum of perturbations 
is attained, in agreement with observations. Indeed, for an 
exponential gauge kinetic function, we obtain $n_s\simeq 0.97-0.98$ 
with negligible running and tensor fraction.

The vector field could contribute to the curvature perturbation
$\zeta$ if it acts as a vector curvaton. We found that the contribution 
to statistical anisotropy in $\zeta$ is important only when the gauge 
coupling constant is unnaturally small. This is because the attractor 
solution is such that the vector field contribution to the energy 
density is rather small. So, a long period of vector field oscillations 
is required after the end of inflation for the vector field to become
significant. This requires a small decay width of the vector field and 
thus an unnaturally small gauge coupling constant.

%%%%%%%%%%%%%%%%%%%%%%%%%%%%%%%%%%%%%%%%%%%%%%%%
%% BACKMATTER
%%%%%%%%%%%%%%%%%%%%%%%%%%%%%%%%%%%%%%%%%%%%%%%%

\begin{theacknowledgments}
This work is supported by the European Union under the Marie Curie 
ITN "UNILHC" PITN-GA-2009-237920.
\end{theacknowledgments}



\bibliographystyle{aipproc}   % if natbib is available


%\bibliography{sample}

 %{\typeout{}
  %\typeout{******************************************}
  %\typeout{** Please run "bibtex \jobname" to optain}
  %\typeout{** the bibliography and then re-run LaTeX}
  %\typeout{** twice to fix the references!}
  %\typeout{******************************************}
  %\typeout{}
 %}

%%%%%%%%%%%%%%%%%%%%%%%%%%%%%%%%%%%%%%%%%%%
%% The following lines show an example how to produce a bibliography
%% without the help of the BibTeX program. This could be used instead
%% of the above.
%%%%%%%%%%%%%%%%%%%%%%%%%%%%%%%%%%%%%%%%%%%

%\endinput

\end{document}